\documentclass[12pt]{article}
\usepackage{graphicx}
\usepackage{authblk}

\newcommand{\pienu}{$\pi^+ \rightarrow \mbox{e}^+ \nu$}

\newcommand{\pimunu}{$\pi^+ \rightarrow \mu^+ \nu$}

\newcommand{\pimue}{$\pi^+ \rightarrow \mu^+ \rightarrow \mbox{e}^+$}
\newcommand{\muenunu}{$\mu^+ \rightarrow \mbox{e}^+ \nu \overline{\nu}$}
\newcommand{\pienuh}{$\pi^+ \rightarrow \mbox{e}^+ \nu_H$}

\def\support{\footnote{email address: mischke@triumf.ca}}

\newenvironment{Abstract}{\begin{quotation}  }{\end{quotation}}
\newenvironment{Presented}{\begin{quotation} \begin{center} 
             PRESENTED AT\end{center} 
      \begin{center}\begin{large}}{\end{large}\end{center} \end{quotation}}
\def\Acknowledgements{  \bigskip \begin{center} \begin{large}
             \bf ACKNOWLEDGEMENTS \end{large}\end{center}}
\title{Improved Search for Heavy Neutrinos and a Test of Lepton Universality in the Decay \pienu}

\author[1]{R.E.~Mischke\support}
\affil[1]{TRIUMF, 4004 Wesbrook Mall, Vancouver, B.C., V6T 2A3, Canada}
\author[2]{A.~Aguilar-Arevalo}
\affil[2]{Inst. de Ciencias Nucl., Univ. Nacional Aut\'onoma de M\'exico, CDMX 04510, M\'exico}

\author[3]{ M.~Aoki}
\affil[3]{Physics Department, Osaka University, Toyonaka, Osaka, 560-0043, Japan}

\author[4]{M.~Blecher}
\affil[4]{Virginia Tech., Blacksburg, VA, 24061, USA}

\author[5]{D.I.~Britton}
\affil[5]{SUPA - School of Physics and Astronomy, Univ. of Glasgow, Glasgow, United Kingdom}

\author[6]{D.~vom~Bruch}
\affil[6]{Dept. of Phys. and Astronomy, Univ. of British Columbia, Vancouver, V6T 1Z1, Canada}

\author[1,6]{D.A.~Bryman}

\author[7]{S.~Chen}
\affil[7]{Department of Engineering Physics, Tsinghua University, Beijing, 100084, China}

\author[8]{J.~Comfort}
\affil[8]{Physics Department, Arizona State University, Tempe, AZ 85287, USA}

\author[6]{S.~Cuen-Rochin}

\author[1,9]{L.~Doria}
\affil[9]{Institut f\"ur Kernphysik, Johannes Gutenberg-Universit\"at Mainz, Johann-Joachim-Becher-Weg 45, D 55128 Mainz, Germany}

\author[1]{P.~Gumplinger}

\author[1,10]{A.~Hussein}
\affil[10]{University of Northern British Columbia, Prince George, B.C., V2N 4Z9, Canada}

\author[11]{Y.~Igarashi}
\affil[11]{KEK, 1-1 Oho, Tsukuba-shi, Ibaraki, Japan}

\author[2]{S.~Ito}
\author[12]{S.~Kettell}
\affil[12]{Brookhaven National Laboratory, Upton, NY, 11973-5000, USA}

\author[1]{L.~Kurchaninov}

\author[12]{L.S.~Littenberg}

\author[6]{C.~Malbrunot}
\author[1]{T.~Numao}

\author[5]{D.~Protopopescu}

\author[1]{A.~Sher}

\author[6]{T.~Sullivan}

\author[1]{D.~Vavilov}

\date{}
\begin{document}
\maketitle
\begin{Abstract}
Two results from the PIENU Experiment are presented reporting a test of lepton universality in pion decay and improved limits on heavy neutrinos coupling to positrons. The status of the full analysis for the \pienu~ branching ratio measurement is summarized.
\end{Abstract}
\begin{Presented}
Thirteenth Conference on the Intersections of Particle and Nuclear Physics (CIPANP2018)\\
Palm Springs CA, U.S.A, May 29 - June 3, 2018
\end{Presented}

\section{Introduction}
The PIENU experiment employs stopped pions to make a precise measurement of the branching ratio for the rare decay \pienu. Recently published results based on a subset of the data gave new limits on the hypothesis of lepton universality \cite{PRL}. 
The experiment is also sensitive to the presence of admixtures of massive neutrinos $\nu_H$ emitted in the decay \pienuh~ for $60 < M_{\nu_H} < 135~\mbox{MeV/c}^2$ and has set improved limits on the coupling of a $\nu_H$ to the positron ($|U_{ei}|^2$) \cite{nuPRD}. The status of the full analysis for the branching-ratio measurement is also discussed.

\section{Experiment}
\begin{figure}[htb]
\centering
\includegraphics[height=2.5in]{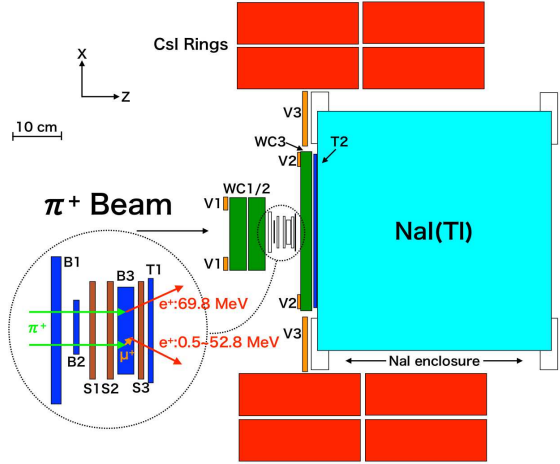}
\caption{Schematic of the PIENU detector.}
\label{fig:detector}
\end{figure}

The PIENU experiment at TRIUMF used a 75 MeV/c pion beam operating with an intensity of 60 kHz and containing 14\% muons and 1\% positrons.  The detector \cite{NIM1} is shown in Fig. \ref{fig:detector}.  Incident pions passed through two wire chambers (WC1/2), two scintillators (B1 and B2), and two silicon strip detectors (S1 and S2) before stopping in an active target (B3). The trigger required detection of an incident pion and an outgoing positron detected by scintillators T1 and T2. Positrons came directly from \pienu~ decay or from the dominant pion decay mode \pimunu~ followed by \muenunu, which is referred to here as \pimue. (The muons were confined to the target where they also decayed.)

The direction of positrons from \pienu~ decay and from the \pimue~ chain was measured by S3 and WC3, and positron energies were measured in a calorimeter (NaI(Tl) plus CsI), which had 2.2\% FWHM energy resolution at 70 MeV. The detector was located in an enclosure used for temperature stabilization. Data taking started in 2009 and was completed in 2012.

\section{Heavy Neutrino Search}
Many extensions of the Standard Model include additional massive neutrinos. The $\nu$MSM includes three sterile neutrinos, two of which may have masses in the range probed by meson decays \cite{Boyarsky}. Other models (such as dark matter or thermalization) also have neutrino masses in the $\mbox{MeV/c}^2$ range \cite{Bertoni}.

\begin{figure}[htb]
\centering
\includegraphics[height=1.6in]{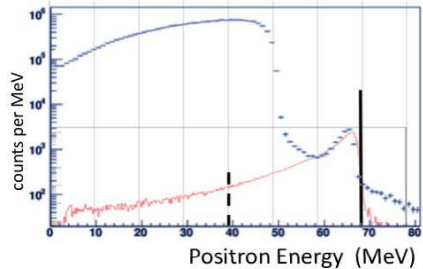}
\caption{Energy spectrum of positrons in the PIENU detector. The vertical black lines correspond to monoenergetic positrons from pion decay (solid for the usual decay with massless neutrinos and dashed for a heavy neutrino mass of $90~\mbox{MeV/c}^2$). The red histogram, from a Monte Carlo simulation, illustrates the effect of the detector response on \pienu~ decay positrons. The blue histogram shows what is actually observed in the time window of 520 ns following the decay, with the dominant feature coming from the $\pi \rightarrow \mu\rightarrow e$ decay chain.}
\label{fig:spectrum}
\end{figure}

The PIENU technique to search for evidence of a heavy neutrino is illustrated in Fig. \ref{fig:spectrum}. The usual \pienu~ decay yields a monoenergetic positron with an energy of 69.8 MeV, while the analagous decay into a heavy neutrino yields a positron with reduced energy. For example, the positron has an energy of approximately 40 MeV if $M_{\nu_H} = 90~\mbox{MeV/c}^2$. In the region of positron energies below 50 MeV, the observed spectrum of positrons including detector response is dominated by positrons from the decay chain \pimue, which complicates the search.

\begin{figure}[htb]
\centering
\includegraphics[height=1.5in]{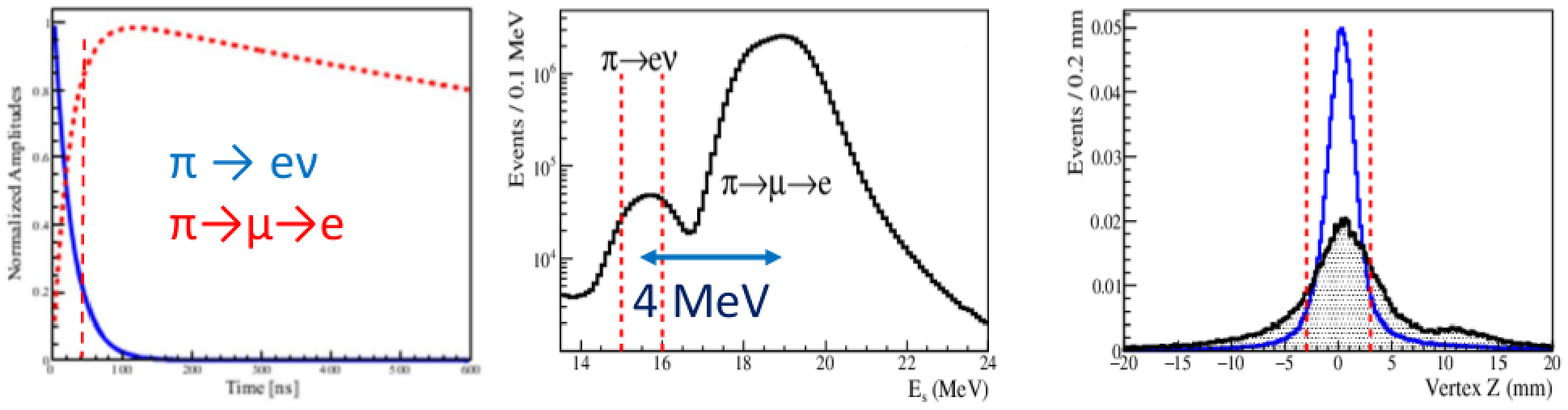}
\caption{\pimue~ suppression cuts include time (left), pion energy (middle), and Z vertex (right). The pion energy ($E_{s}$) is the sum measured in B1, B2, S1, S2, and B3. The Z vertex histograms are for events with positron energy $E_{e^+}>52$~MeV and $E_{e^+}<52$~MeV (shaded histogram). The cuts applied are indicated by the red vertical dashed lines.}
\label{fig:suppression}
\end{figure}

\begin{figure}[htb]
\centering
\includegraphics[height=1.4in]{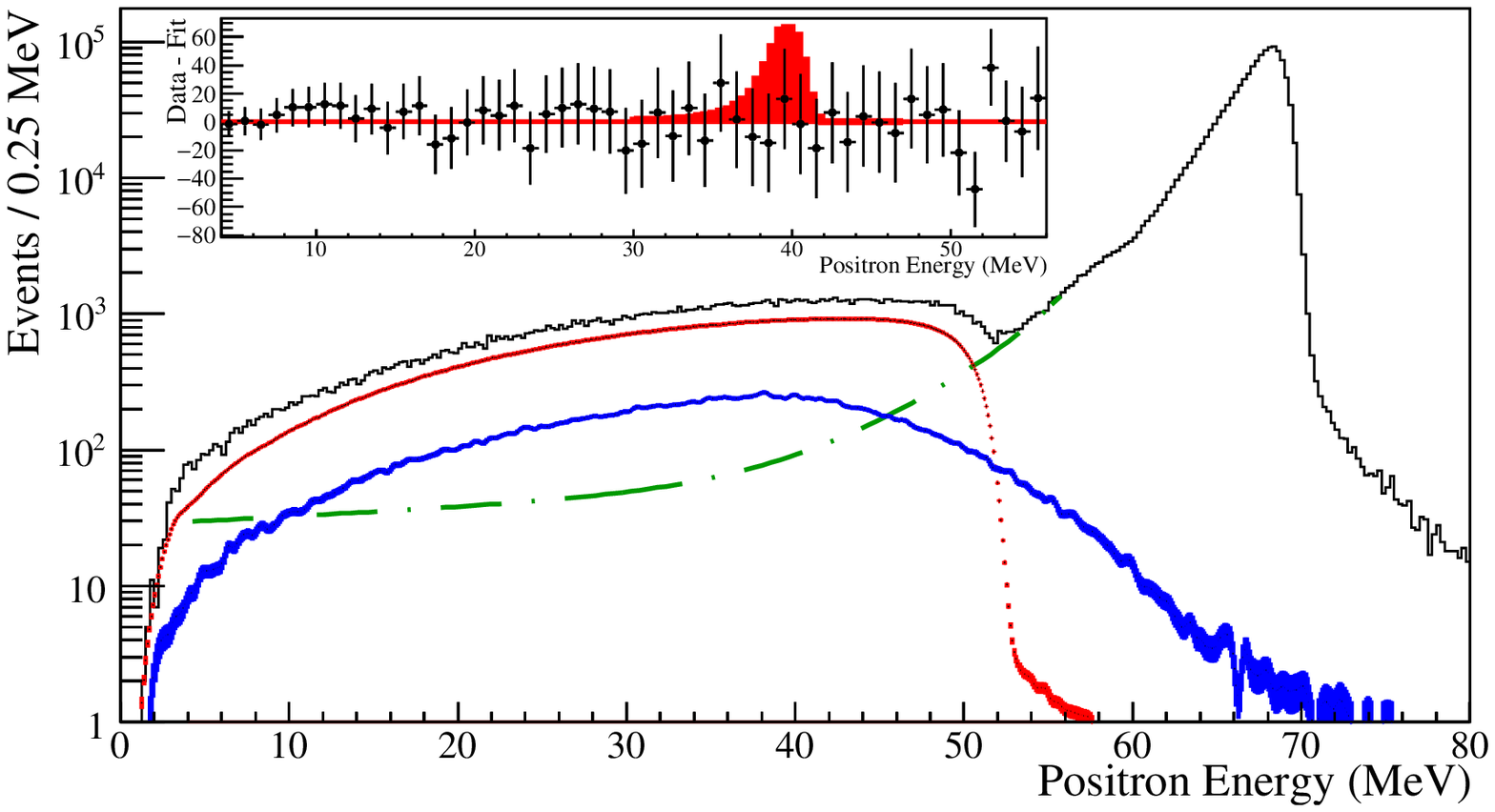}
\includegraphics[height=1.4in]{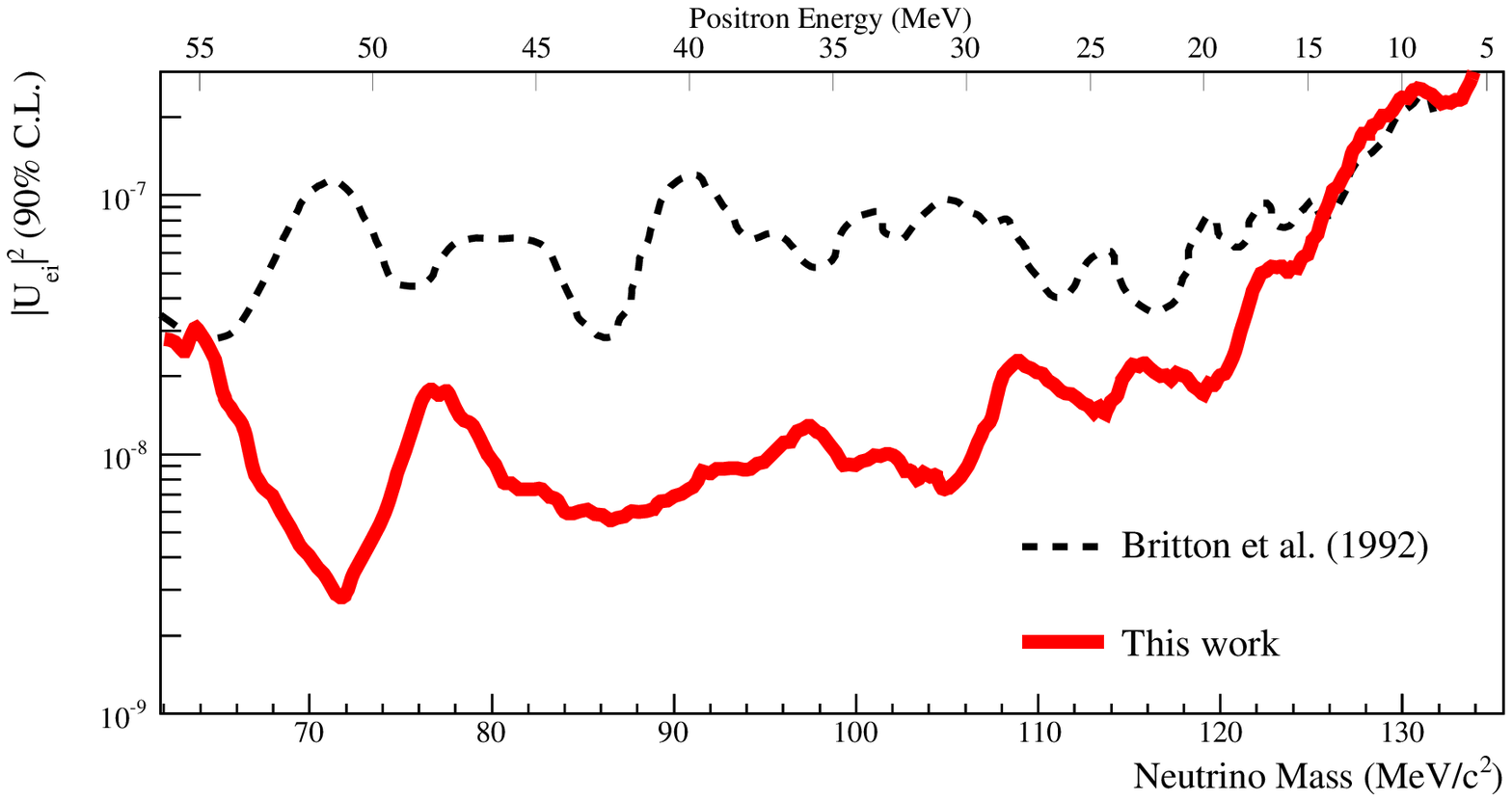}
\caption{Left: Background-suppressed energy spectrum (black histogram). Fitted components include muon decays in flight (blue line, from MC), \pienu~ (green, dot-dashed line, fit to MC), and \pimue~ (red dashed line, from late-time data events). The insert shows the (rebinned) residuals (Data--Fit) for the background fit including a sample signal at 40 MeV for $|U_{ei}|^2 = 10^{-8}$. Right: 90\% upper limits on mixing $|U_{ei}|^2$ of heavy neutrinos coupled to positrons (thick red line). The black dashed line shows the results from~\cite{oldneutrino}.}
\label{fig:shapes_fit}
\end{figure}

To improve the sensitivity for finding a monoenergetic positron under the background, several cuts are applied to suppress the background. As shown in Fig. \ref{fig:suppression}, these cuts involve timing, target energy, and Z vertex (decay vertex position along the beam direction obtained from pion and positron tracks) measurements.  The resulting spectrum is shown in the left panel of Fig. \ref{fig:shapes_fit} as the black histogram. This suppressed spectrum is fit with known shapes to describe its components. These are also shown in Fig. \ref{fig:shapes_fit} and include the extrapolated tail of the \pienu~peak, the \pimue~ shape taken from late decays, and the muon decay-in-flight shape from Monte Carlo (MC) simulations. The insert shows the residuals of the fit, and an example of a heavy neutrino shape at 40 MeV for $|U_{ei}|^2 = 10^{-8}$.

The results for the upper limits as a function of heavy neutrino mass are shown in the right panel of Fig. \ref{fig:shapes_fit} from \cite{nuPRD}. These results represent up to an order of magnitude improvement over previous limits~\cite{oldneutrino}.

\section{\pienu~Branching Ratio}

The theoretical prediction for the \pienu~branching ratio, assuming lepton universality and including radiative and structure corrections \cite{Berman}, is

\begin{equation}
R^{th}_{e/\mu} = \frac{\Gamma_{\pi \rightarrow e \nu_e}}{\Gamma_{\pi \rightarrow \mu \nu_{\mu}}} = 1.2352(2) \times 10^{-4}.
\end{equation}

\begin{figure}[htb]
\centering
\includegraphics[height=1.8in]{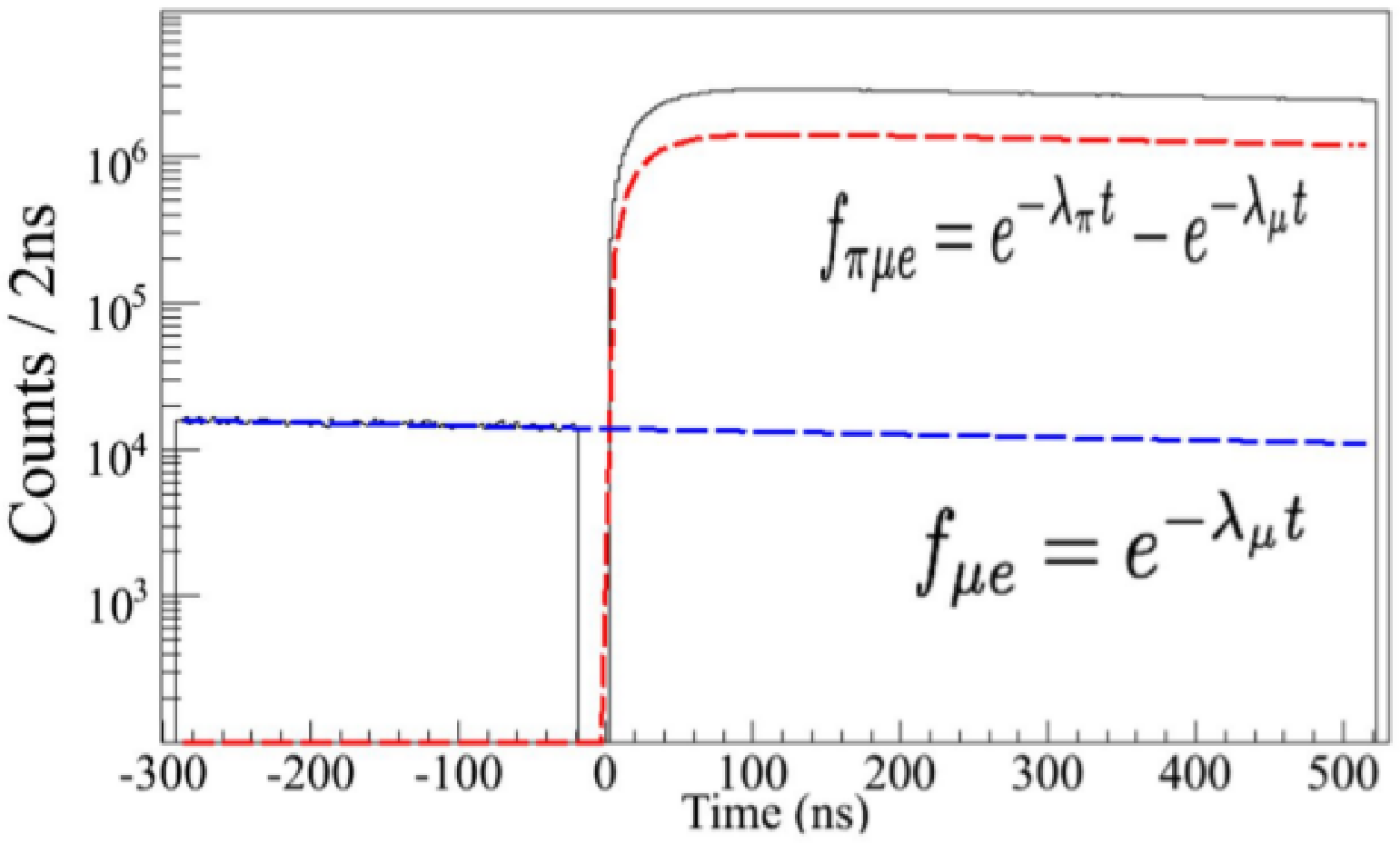}
\includegraphics[height=1.8in]{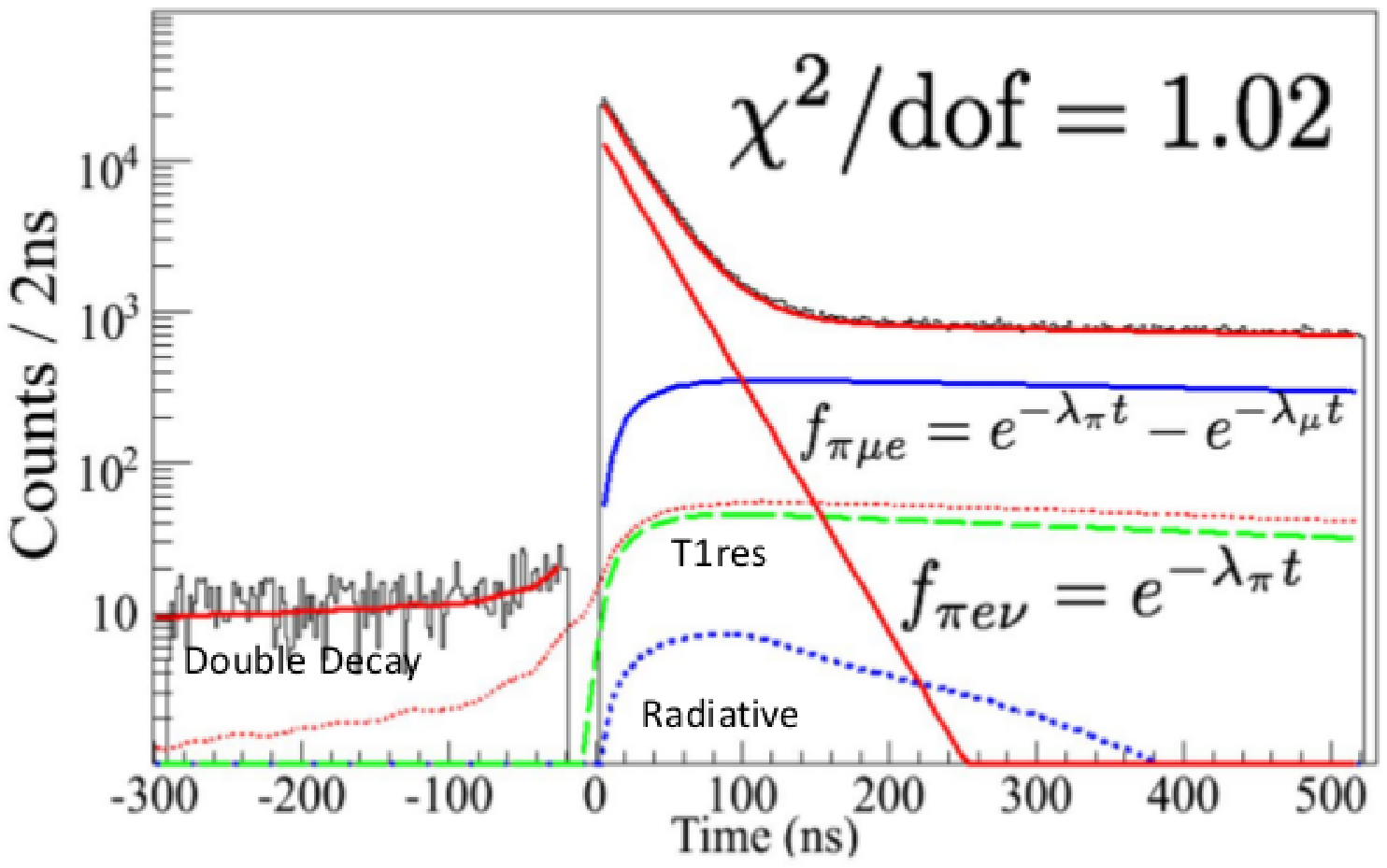}
\caption{Time spectra with fit components for low-energy (left) and high-energy (right) portions \cite{PRL}.}
\label{fig:time}
\end{figure}

To compare with this prediction, and to test lepton universality or other deviations from the Standard Model, the observed energy spectrum of positrons (see Fig. \ref{fig:spectrum}) was divided into low- and high-energy portions with a cut at 52 MeV. The time spectrum of events thus separated is shown in Fig. \ref{fig:time}. These spectra are simultaneously fitted to known component shapes to extract the branching ratio.

The fit yielded the raw branching ratio. Some corrections must be included. The dominant correction was for the events from \pienu~decay with energies below the cut value. This low-energy tail was evaluated from special data taken with a monoenergetic positron beam. The result for the branching ratio, based on 12\% of the data, is presented in Table \ref{table:result}.

\begin{table}[thb]
\centering
\vspace*{0.5cm}
\begin{tabular}{l|rrr}
\hline
\hline
 & Values & \multicolumn{2}{c}{Uncertainties} \\
 & &\hspace*{0.5cm} $Stat$ &\hspace*{0.5cm} $Syst$\\
\hline
$R_{e/\mu}^{Raw}$ $(10^{-4})$ & 1.1972& 0.0022&  0.0005\\
\hspace{0.5cm} $\pi$,$\mu$ lifetimes & & & 0.0001\\
\hspace{0.5 cm} Other parameters & & & 0.0003\\
\hspace{0.5 cm} Excluded components & & & 0.0005\\
\hline
Corrections\\
\hspace{0.5 cm} Acceptance & 0.9991 & & 0.0003\\
\hspace{0.5 cm} Low-energy tail & 1.0316& & 0.0012\\
\hspace{0.5 cm} Other & 1.0004& & 0.0008\\
\hline
$R_{e/\mu}^{Exp}$ $(10^{-4})$ & 1.2344& 0.0023& 0.0019\\
\hline
\hline
\end{tabular}
\caption{The table includes the raw branching ratio with its 
statistical and systematic uncertainties, the multiplicative 
corrections with their errors, and the result after applying 
corrections.
}
\label{table:result}
\end{table}

The ratio between the experimental and theoretical branching ratios provides a test of lepton universality:

\begin{equation}
\frac{g_e}{g_{\mu}} = \sqrt \frac{R^{exp}_{e/\mu}}{R^{th}_{e/\mu}} = 0.9996 \pm 0.0012,
\end{equation}
where $g_e$ and $g_\mu$ are possibly distinct weak interaction coupling factors for e and $\mu$.

\begin{figure}[htb]
\centering
\includegraphics[height=3.0in]{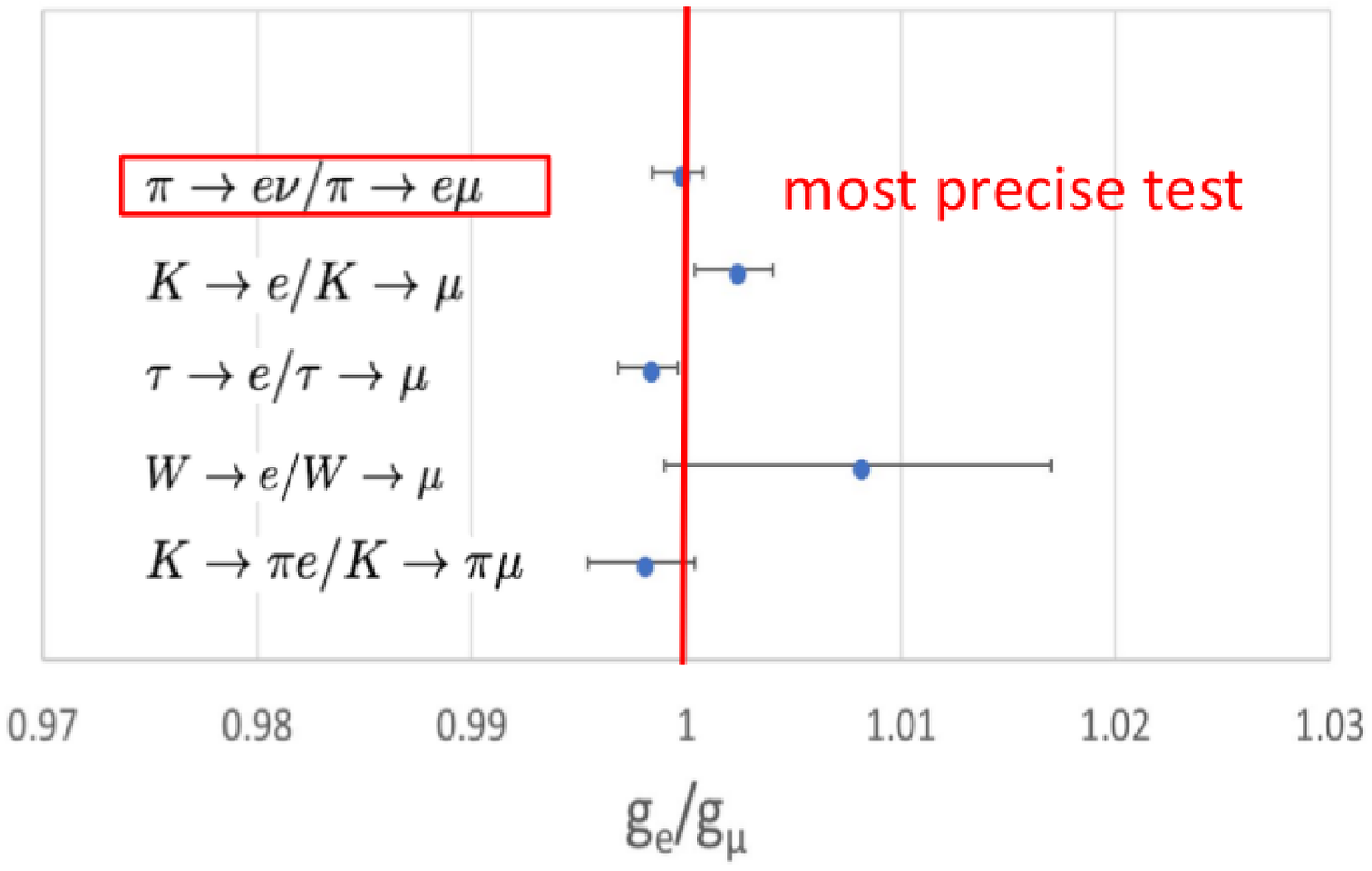}
\caption{Tests of lepton universality.}
\label{fig:LU}
\end{figure}

This result, together with similar tests of lepton universality, is shown in Fig. \ref{fig:LU}. These tests are complementary to tests with $\tau$ decay and with heavy quarks, where the ensemble of B decay results suggest possible violations of lepton universality \cite{CERN}.

The analysis of the full dataset for PIENU is in the final stages. The dataset processed is about 8 times larger than for the 2015 result with about $10^7$ \pienu~ events. The value of the branching ratio will remain blinded until all cuts and systematic uncertainties have been finalized. The expected statistical uncertainty will be close to the design value of 0.1\% for the branching ratio. It is dependent on the choice of cut for the solid angle, which will be an acceptance of about 20\%, chosen to minimize the total uncertainty.

\Acknowledgements
This work was supported by the Natural Sciences and Engineering Research Council of Canada and TRIUMF through a contribution
from the National Research Council of Canada, and by Research Fund for Doctoral Program of Higher Education of China, and
partially supported by KAKENHI (18540274, 21340059) in Japan.
One of the authors (M.B.) was supported by US National Science Foundation Grant Phy-0553611.
We are indebted to Brookhaven National Laboratory for the loan of the crystals.
We would like to thank the TRIUMF detector, electronics and DAQ groups for the extensive support.

\end{document}